\documentclass[journal=jctcce, manuscript=article, layout=twocolumn]{achemso}
\setkeys{acs}{articletitle=true}

\usepackage{chemformula} 
\usepackage[T1]{fontenc} 
\usepackage{amsmath, amsfonts}
\usepackage{physics}
\usepackage{comment}
\usepackage{booktabs}
\usepackage{multirow}
\usepackage{graphicx}
\usepackage{textcomp}  
\mciteErrorOnUnknownfalse
\usepackage{subfigure}
\usepackage{mhchem}
\usepackage{multicol}
\usepackage{mathtools}

\newcommand*\supp[1]{\, \mathrm{supp}\bigl(#1\bigr)}
\DeclareMathOperator*{\argmax}{arg\, max}
\DeclareMathOperator*{\argmin}{arg\, min}

\newcommand{\fnorm}[1]{\norm{ #1 }_{\mathrm{F}}}

\newcommand*\Hop{\widehat{H}}

\newcommand*\bigO{O}

\newcommand*\fname{coordinate descent FCI}
\newcommand*\Fname{Coordinate descent FCI}
\newcommand*\sname{CDFCI}
\newcommand*\xsname{xCDFCI}

\newcommand*\bbR{\mathbb{R}}
\newcommand*\calI{\mathcal{I}}

\newcommand*\itell{{(\ell)}}
\newcommand*\itellp{{(\ell+1)}}

\newcommand*\NFCI{N_{\text{FCI}}}
\newcommand*\hset[1]{\calI_H(#1)}

\author{Zhe Wang}
\affiliation[Duke]{Department of Mathematics, Duke University}
\author{Zhiyuan Zhang}
\affiliation[USTC]{School of Future Technology, 
University of Science and Technology of China}
\author{Jianfeng Lu}
\email{jianfeng@math.duke.edu}
\affiliation[Duke]{Department of Mathematics, Duke University}
\alsoaffiliation{Department of Chemistry and Department of Physics,
Duke University}
\author{Yingzhou Li}
\affiliation[Fudan]{School of Mathematical Sciences, Fudan University}
\email{yingzhouli@fudan.edu.cn}

\title[Coordinate Descent FCI Excited State]
{Coordinate Descent Full Configuration Interaction for Excited States}

\abbreviations{FCI, CDM, \sname{}}

\begin{document}

%
%
%
%
%

\begin{abstract}
An efficient excited state method, named \xsname{}, in the configuration
interaction framework, is proposed. \xsname{} extends the unconstrained
nonconvex optimization problem in coordinate descent full configuration
interaction~(\sname{}) to a multicolumn version, for low-lying excited
states computation. The optimization problem is addressed via a tailored
coordinate descent method. In each iteration, a determinant is selected
based on an approximated gradient, and coefficients of all states
associated with the selected determinant are updated. A deterministic
compression is applied to limit memory usage. We test \xsname{} applied to
\ch{H2O} and \ch{N2} molecules under the cc-pVDZ basis set. For both
systems, five low-lying excited states in the same symmetry sector are
calculated together with the ground state. \xsname{} also produces
accurate binding curves of carbon dimer in the cc-pVDZ basis with chemical
accuracy, where the ground state and four excited states in the same
symmetry sector are benchmarked.
\end{abstract}

\section{Introduction}

Excited state computations are of great importance in understanding and
predicting many phenomena in photochemistry, spectroscopy, and others.
Compared to the ground state computation, excited state computations are
more challenging for wavefunction ansatz based methods, including
Hartree-Fock methods~\cite{Morokuma1972, Barca2014}, configuration
interaction methods~\cite{Sherrill1995}, and coupled cluster
methods~\cite{Geertsen1989, Stanton1993, Nooijen1997}, etc. The excited
states in general have multi-reference characters, and the wavefunction
ansatzes in these methods limit the representation of dynamic
correlations. Similarly, density functional theory (DFT)
methods~\cite{Onida2002, Rocca2012, Yang2014, Lu2017c, Hu2020} and
time-dependent DFT methods~\cite{Runge1984, Burke2005} find it more
challenging to calculate the excited states than the ground state.

Under the full configuration interaction~(FCI) framework, it is also
considered more challenging to calculate the excited states but the
difficulty is not as severe as the aforementioned methods. In general,
there are two types of challenges for excited state computations under
FCI. First, due to the natural multi-reference features of excited states,
the discretization basis set should be of larger sizes than that in ground
state computation and the corresponding FCI matrix size should be larger.
Second, the energy gaps between excited states are in general smaller than
that between the ground state and the first excited state, which would
lead to more iterations in iterative eigensolvers before converging. In
this paper, we propose \xsname{} for excited state computation under the
FCI framework. The method is closely related to the recently developed
efficient FCI solver, by three of the authors,
\fname{}~(\sname{})\cite{Wang2019}.

Many modern FCI solvers have been developed for ground state computation
in the past two decades, together with their extensions to excited state
computations. Density matrix renormalization group~(DMRG)~\cite{White1992,
Chan2011, Schollwock2011, Baiardi2020} uses matrix product state as the
wavefunction ansatz and applies an iterative sweeping procedure as an
eigensolver. Various strategies~\cite{Chandross1999, Sharma2015} are
proposed to address excited states one by one. FCI quantum Monte
Carlo~(FCIQMC) and its variants~\cite{Booth2009, Cleland2010,
Petruzielo2012} use quantum Monte Carlo walker idea to reduce the
computational cost. In its extension to excited state
computations~\cite{Blunt2015}, several groups of walkers are used to
represent excited states, and an orthogonal projection is introduced
between iterations to prevent groups from collapsing into the ground
state. Selected-CI is a group of FCI solvers based on sequential
configuration selections, including adaptive configuration
interaction~(ACI)~\cite{Schriber2016}, heat-bath configuration
interaction~(HCI)~\cite{Holmes2016, Li2018b, Sharma2017}, and adaptive
sampling configuration interaction~(ASCI)~\cite{Tubman2016}. Extending
selected-CI methods to excited state computations is straightforward.
After a small modification of the selection criteria~\cite{Tubman2016,
Holmes2017, Schriber2017}, the excited states are computed by solving the
low-lying eigenstates of the reduced Hamiltonian matrix. FCI fast random
iteration~(FCI-FRI)~\cite{Greene2019a} adopts a bias-free sampling
procedure to compress the wavefunction under the power method framework.
In the excited state version of FCI-FRI~\cite{Greene2022}, the iterative
method is a multicolumn version power method, where the normalization is
carried out every iteration and the orthogonalization is carried out every
few iterations. Recent review papers~\cite{RN1084, RN1159} summarize other
FCI-related methods as well. \Fname{}~\cite{Wang2019} reformulates the
eigenvalue problem as an unconstrained optimization problem, which is the
single-column version of the optimization problem used in this
paper~\eqref{eq:opt}. Then a coordinate-descent method is applied to
address the optimization problem, where the coordinates are selected based
on the magnitude of the gradient vector and the stepsize is calculated
from an exact linesearch. Importantly, a tailored compression strategy is
applied to limit the growth of nonzeros in the state vector and, hence,
limit memory usage. The compression is not applied to the state vector $c$
directly. Instead, it is applied to $b = Hc$ for $H$ being the Hamiltonian
matrix to truncate small updates that increase the memory cost. The
compression in \Fname{} is carefully designed so that the Rayleigh
quotient could be accurately evaluated and the second-order energy
estimation becomes available.

Moreover, FCI problems have also attracted attention from the numerical
linear algebra community in recent years. Many algorithms and
analyses~\cite{Lim2017, Li2019c, Lu2020, Hernandez2019, Gao2022, Gao2023a,
Gao2023b} influence the developments above. Other works attempt to
incorporate machine learning and reinforcement learning technique to
accelerate the FCI calculation~\cite{Coe2019, Goings2021}.

In this paper, we extend \sname{} to excited state computations and name
the method as \xsname{}. The unconstrained optimization problem in
\sname{} is extended to a multi-column version to accommodate low-lying
excited states. The coordinate-descent method used to optimize the
objective function is replaced by a row-block descent scheme in \xsname{}
and the compression is still carried out in an entrywise way. The
multi-column vector in \xsname{} does not converge to the ground state and
low-lying excited states directly. Instead, it converges to a subspace
formed by the ground and low-lying excited states. The eigenvectors can be
recovered by a post-processing procedure. Most importantly, all desired
features of the original \sname{} are preserved. Symmetries, including
time-reversal symmetry and angular momentum symmetry, are implemented to
reduce both computational and memory costs when the computation is
restricted to a symmetry sector. Finally, numerical results on \ch{H2O},
\ch{N2} are included to demonstrate the efficiency of \xsname{} for
excited state computations. We also report the binding curve of \ch{C2}
obtained using \xsname{} for singlet. 

The rest of the paper is organized as follows. Section~\ref{sec:cdfci}
introduces \xsname{} for excited state computations and other related
discussions. Section~\ref{sec:num} provides numerical examples of
\xsname{}. The paper is concluded in Section~\ref{sec:conclusion}.

\section{\xsname{}}
\label{sec:cdfci} 

We introduce \xsname{} in this section and discuss some implementation
details. Notations are kept the same as that in~\citet{Wang2019} as much
as possible. In the following, we first propose the unconstrained
optimization problem for excited state computations, then explain the
\xsname{} algorithm step-by-step, and finally discuss its implementation
details: initialization, stopping criteria, and symmetry. 

\subsection{Optimization formula for excited state computations}

Given a spin-orbital set $\{\chi_p\}$, we denote the creation and
annihilation operator as $\hat{a}_p^\dagger$ and $\hat{a}_q$ respectively.
The Hamiltonian operator, under the second quantization, is given by  
\begin{equation}
    \Hop = \sum_{p,q} t_{pq} \hat{a}_p^\dagger \hat{a}_q 
    + \sum_{p,r,q,s} v_{prqs} \hat{a}_p^\dagger \hat{a}_r^\dagger
    \hat{a}_s \hat{a}_q,
\end{equation}
where $t_{pq}$ and $v_{prqs}$ are one-body and two-body integrals
respectively. The $K$ low-lying states of the time-independent
Schr{\"o}dinger equation can be obtained by solving, 
\begin{equation} \label{eq:excitedstates}
    \Hop \ket{\Phi_k} = E_k \ket{\Phi_k}
\end{equation}
for $k = 0, 1, \dots, K-1$, where $E_0$ is the smallest eigenvalue
associated with the ground state $\ket{\Phi_0}$, $E_1$ is the second
smallest eigenvalue associated with the first excited state
$\ket{\Phi_1}$, and so on, $\{\ket{\Phi_k}\}_{k=0}^{K-1}$ are orthogonal
to each other.\footnote{With some abuse of terminology, we will also refer
the ground state as the $0$-th excited state when it is convenient to do
so.} Throughout this paper, we assume that all $E_0, E_1, \dots, E_{K-1}$
are negative. This assumption can be made without loss of generality, as
otherwise, we can shift the Hamiltonian by a constant. We further denote
the Slater determinants as $\{\ket{D_i}\}_{i=1}^N$ for $N = \NFCI$ being
the size of the entire electron-preserving configuration space. Using
$\{\ket{D_i}\}_{i=1}^N$ as the basis, the ground state and excited states
are discretized as,
\begin{equation} \label{eq:eigenstates}
    \ket{\Phi_k} = \sum_i V_{i,k} \ket{D_i},
\end{equation}
and coefficients $V_{i,k}$ forms a matrix $V$ of size $N \times K$
satisfying the orthonormality constraint, $V^\top V = I$ for $I$ being an
identity matrix of size $K \times K$. The Hamiltonian operator is
discretized as the Hamiltonian matrix $H$ with its $(i,j)$-th entry being
$H_{ij} = \mel**{D_i}{\Hop}{D_j}$. After the discretization,
solving~\eqref{eq:excitedstates} is reduced to solving for the low-lying
$K$ eigenpairs of $H$, where the major computational difficulty comes from
the factorial scaling of $\NFCI$ with respect to the number of
spin-orbitals and electrons.

Now we extend the unconstrained optimization problem in
\sname{}~\cite{Wang2019} to excited states. The optimization problem is
extended as, 
\begin{equation} \label{eq:opt}
    \min_{C \in \bbR^{N \times K}} f(C),
\end{equation}
for
\begin{equation}
    f(C) = \fnorm{H + C C^\top }^2,
\end{equation}
where $C$ is a matrix of size $N$ by $K$. When $K = 1$, \eqref{eq:opt} is
the same as the optimization problem in \citet{Wang2019} The gradient of
$f(C)$ admits, 
\begin{equation} \label{eq:grad}
    G = \nabla f = 4 H C + 4 C \big( C^\top C \big).
\end{equation}
As has been analyzed in \citet{Gao2022}, the unconstrained optimization
problem~\eqref{eq:opt} has many stationary points, but has no spurious
local minima. All local minima are global minima of the form, 
\begin{equation} \label{eq:globalminima}
    V \sqrt{-\Lambda} Q,
\end{equation}
where $\Lambda \in \bbR^{K \times K}$ is a diagonal matrix with its
diagonal entries being $E_0, E_1, \dots, E_{K-1}$, $V \in \bbR^{N \times
K}$ is the corresponding eigenvector matrix as defined in
\eqref{eq:eigenstates}, and $Q \in \bbR^{K \times K}$ is an arbitrary
orthogonal matrix such that $Q^\top Q = Q Q^\top = I$.

Generally, gradient-based first-order methods, including the coordinate
descent method, avoid saddle points and converge to a global minimum
almost surely~\cite{Chen2021}. We remark that the minimizers of
\eqref{eq:opt} only give the eigenspace due to the arbitrary $Q$ in
\eqref{eq:globalminima}. To get eigenvectors, we need a post-processing
step to retrieve eigenvectors when needed. The post-processing part is
computationally cheap and costs no additional memory.

\subsection{Algorithm}

The algorithm we propose for excited state computations is a coordinate
descent method applying to \eqref{eq:opt}, where some specifics are
designed to fully incorporate the properties of FCI problems. We introduce
our algorithm step by step. Throughout the algorithm, two matrices $C$ and
$B$ are kept: $C$ is the iterator targeting \eqref{eq:globalminima} and
$B$ is used to track $HC$, \latin{i.e.}, $B \approx HC$.  Further, we use
superscript in the parenthesis to denote iteration index, \latin{e.g.},
$C^\itell$ denotes the iterator at the $\ell$-th iteration. Colon notation
is used to denote the entire row or column, \latin{e.g.}, $C^\itell_{i,:}$
denotes the $i$-th row of $C^\itell$.

The \xsname{} algorithm is composed of an iterative part with 5 steps and
a post-processing step. At each iteration,  the first step selects a
determinant with maximum absolute value in an approximated gradient of
\eqref{eq:opt}. The second step then conducts a linesearch and updates
$C$, where a fourth-order polynomial is minimized to determine the optimal
stepsize. In the third and fourth steps, the corresponding update to $B$
is calculated with compression and a row of $B$ is recalculated to improve
accuracy with minimal additional cost. In the last step of the iterative
part, energies are estimated via a generalized Rayleigh quotient
procedure. When the iteration converges according to some stopping
criteria, a post-processing step could be carried out to obtain the ground
state vector and excited state vectors. In the following, we will explain
each step of \xsname{} in detail.

\subsubsection*{Step 1: Determinant select}

This step aims to select a determinant for the update, which potentially
leads to greatest decay in $f(C)$. The determinant selection strategy is
as follows, 
\begin{equation} \label{eq:detselect}
    i^\itellp = \argmax_{\substack{j \in \hset{i^\itell} \\ 0 \leq k < K}}
    \abs{4 B^\itell_{j,k} + 4 C^\itell_{j,:} \Big[\big( C^\itell
    \big)^\top C^\itell \Big]_{:,k}},
\end{equation}
where $i^\itellp$ is the argument $j$ achieving the maximum value. Here
$\hset{i^\itell}$ denotes the set of determinants connected to $i^\itell$
via $H$, \latin{i.e.}, for any $j \in \hset{i^\itell}$, $H_{i^\itell j}$
is nonzero and for any $j \not\in \hset{i^\itell}$, $H_{i^\itell j}$ is
zero. Due to the existence of zeros in one- and two-body integrals,
$\hset{i^\itell}$ is a subset of the single and double excitations from
the $i^\itell$ determinant. The intuition behind~\eqref{eq:detselect} is
related to the gradient of $f(C)$~\eqref{eq:grad}. Comparing
\eqref{eq:detselect} and \eqref{eq:grad}, we notice that the determinant
is selected to be the row containing the absolutely largest gradient
entry, so that it potentially leads to the greatest reduction of the
objective function.

\subsubsection*{Step 2: Coefficient update}

Given a selected determinant $i^\itellp$, we seek the best stepsize $\tau$
and move the $i^\itellp$-th row of the coefficient matrix $C^\itell$ along
the gradient direction with the stepsize. The best stepsize $\tau$ is
achieved via solving, 
\begin{equation} \label{eq:linesearch}
    \tau = \argmin_{\tilde{\tau}} f\big( C^\itell + \tilde{\tau}
    e_{i^\itellp} \widetilde{G}_{i^\itellp,:} \big),
\end{equation}
where $e_{i^\itellp}$ is a vector with $i^\itellp$-th entry being one and
zero otherwise, and
\begin{equation}
    \widetilde{G}_{i^\itellp,:} = 4 B^\itell_{i^\itellp,:} + 4
    C^\itell_{i^\itellp,:} \big( C^\itell \big)^\top C^\itell
\end{equation}
is the $i^\itellp$-th row of the approximated gradient \eqref{eq:grad}.
Solving \eqref{eq:linesearch} is actually minimizing a fourth-order
polynomial of $\tilde{\tau}$ and all polynomial coefficients can be
evaluated in $\bigO(K^2)$ operations (details can be found in
Appendix~\ref{app:linesearch}). Once the stepsize $\tau$ is determined, we
update $C^\itell$ as follows,
\begin{equation}
    C^\itellp_{i,:} =
    \begin{cases}
        C^\itell_{i,:} + \tau \widetilde{G}_{i,:} & \text{if } i =
        i^\itellp; \\
        C^\itell_{i,:} & \text{otherwise}.
    \end{cases}
\end{equation}

\subsubsection*{Step 3: Coefficient compression}

Throughout the algorithm, we keep all entries of $C$. While, for $B = HC$
without compression, the number of nonzeros in $HC$ is much larger than
that in $C$. We cannot afford to store $HC$ in memory. Hence, we compress
the representation of $B$.

We use $\supp{B}$ to denote the set of determinants containing at least
one nonzero coefficient, \latin{i.e.}, $\supp{B} = \{i: \max_k
\abs{B_{i,k}} > 0\}$. Then we update and compress $B^\itell$ as follows,
for $i = i^\itellp$,
\begin{equation} \label{eq:Bupdate}
    B^\itellp_{j,:} =
    \begin{cases}
        B^\itell_{j,:} + \tau H_{j,i} \widetilde{G}_{i,:} & \text{if
        } j \in \supp{B^\itell} \\
        \tau H_{j,i} \widetilde{G}_{i,:} &
        \substack{
            \text{if } j \not\in \supp{B^\itell} \text{ and } \\
            \max_k \abs{\tau H_{j,i} \widetilde{G}_{i,k}}
            > \varepsilon}
    \end{cases},
\end{equation}
where $\varepsilon$ is the pre-defined compression threshold.
Equation~\eqref{eq:Bupdate} indicates that: for all pre-existing
determinants in $B$, the coefficients are updated accurately; while for
new determinants, the coefficients are added only if they contain an
important update. Obviously, the compression limits the growth of nonzeros
in $B$, and thus the data storage cost.

Now we explain the indirect connection to the compression of $C$.
According to \eqref{eq:detselect}, when a determinant is not in
$\supp{B^\itell}$, the corresponding gradient is zero, hence the
determinant will not be selected, which in turn limits the growth of
nonzeros in $C$. Therefore, all compressions are explicitly applied to $B$
only, indirectly limiting the growth of nonzeros in $C$.

\subsubsection*{Step 4: Coefficient recalculation}

In \eqref{eq:Bupdate}, we already compute all nonzero entries in the
$i^\itellp$-th column of $H$. Now we reuse these results to refine
coefficients in $B$. The $i^\itellp$-th row in $B$ is recalculated as
follows, 
\begin{equation}
    \begin{split}
        B^\itellp_{i^\itellp,:} = & \sum_{j \in \hset{i^\itellp}}
        H_{i^\itellp,j} C^\itellp_{j,:} \\
        = & \sum_{j \in \hset{i^\itellp}}
        H_{j,i^\itellp} C^\itellp_{j,:},
    \end{split}
\end{equation}
where the second equality is due to the symmetry property of the
Hamiltonian.~\footnote{If the Hamiltonian matrix is complex Hermitian,
then a complex conjugate is needed in the equation.} This recalculation of
$B^\itellp_{i^\itellp,:}$ is of essential importance when the
$i^\itellp$-th determinant is added to $C^\itell$ for the first time. It
removes potential errors made by compressions from earlier iterations and,
together with \eqref{eq:Bupdate}, keeps $B_{i^\itellp,:} \equiv
H_{i^\itellp,:}C$ for all later iterations. From a numerical analysis
viewpoint, the recalculation also preserves numerical accuracy. Since the
number of iterations in \xsname{} could easily go beyond $10^8 - 10^{10}$,
the accumulation of the numerical error caused by the finite precision
computations in the worst case grows linearly with respect to the number
of operations and would destroy the accuracy of energies. Regularly
recalculating $B^\itellp_{i^\itellp,:}$ keeps $B_{i^\itellp,:} \equiv
H_{i^\itellp,:}C$ at a low level of numerical error.

\subsubsection*{Step 5: Energy estimation}

Given a coefficient matrix $C^\itellp$, the energy estimation is conducted
through a generalized Rayleigh quotient of second-order accuracy, which
solves a generalized eigenvalue problem of matrix pair $\Big( \big(
C^\itellp \big)^\top H C^\itellp, \big( C^\itellp \big)^\top C^\itellp
\Big)$, \latin{i.e.},
\begin{equation} \label{eq:geneig}
    \Big( \big( C^\itellp \big)^\top H C^\itellp \Big) U =
    \Big(
    \big( C^\itellp \big)^\top C^\itellp \Big)
    U \Gamma,
\end{equation}
for $U$ being eigenvectors and $\Gamma$ being the eigenvalue
matrix~\footnote{We assume $U$ is a $ \bigl( ( C^\itellp )^\top C^\itellp
\bigr) $ orthonormalized eigenvector matrix, \latin{i.e.}, $U^\top \bigl(
( C^\itellp )^\top C^\itellp \bigr) U = I $.}. A detailed discussion on
the accuracy of the Rayleigh quotient refers to
Appendix~\ref{app:rayleigh}. Since only the coefficients of a determinant
are updated, both matrices can be updated accordingly, 
\begin{equation}
    \begin{split}
        \big( C^\itellp \big)^\top C^\itellp = \big( C^\itell
        \big)^\top C^\itell \\
        + \tau \Big( \big(C^\itell_{i^\itellp,:}\big)^\top
        \widetilde{G}_{i^\itellp,:} + \widetilde{G}_{i^\itellp,:}^\top
        C^\itell_{i^\itellp,:}\Big)\\
        + \tau^2 \widetilde{G}_{i^\itellp,:}^\top
        \widetilde{G}_{i^\itellp,:},
    \end{split}
\end{equation}
and,
\begin{equation}
    \begin{split}
        \big( C^\itellp \big)^\top H C^\itellp = \big( C^\itell
        \big)^\top H C^\itell \\
        + \tau \Big( \big(B^\itellp_{i^\itellp,:}\big)^\top
        \widetilde{G}_{i^\itellp,:} + \widetilde{G}_{i^\itellp,:}^\top
        B^\itellp_{i^\itellp,:}\Big)\\
        - \tau^2 H_{i^\itellp i^\itellp} \widetilde{G}_{i^\itellp,:}^\top
        \widetilde{G}_{i^\itellp,:}.
    \end{split}
\end{equation}
Since $B^\itellp_{i^\itellp,:}$ was recalculated in the previous step,
both matrices are numerically accurate and not affected by our
compression. The updated matrix $\big( C^\itellp \big)^\top C^\itellp$ is
also involved and reused in the gradient computation of the next
iteration. After the energy estimation, we check the stopping criteria. If
the criteria are satisfied, we move on to post-processing; otherwise, we
go back to the first step.

\subsubsection*{Post-processing}

When the algorithm converges, energies of low-lying excited states are
already available in $\Gamma$. If excited states are needed for the down
stream tasks, \latin{e.g.}, reduced density matrix computations, the
coefficient matrix $C$ needs to be transformed back to eigenvectors $V$
and the transformation is as simple as,
\begin{equation} \label{eq:VCU}
    V \approx CU,
\end{equation}
where $U$ is the eigenvector matrix in \eqref{eq:geneig}.

\subsection{Implementation}

We now discuss some implementation details, including the data structure
of $C$ and $B$, the stopping criteria, and the symmetry of molecular
systems in the following.

\subsubsection*{Data structure}

In \citet{Wang2019} several data structures have been implemented and
discussed, including the hash table, black-red tree, etc. Among these data
structures, the hash table is the one achieving the best computational
performance for \sname{}. Thus for \xsname{}, we also adopt hash tables as
our overall data structure. For the single-threaded version of our
implementation, a Robin Hood hash table is
adopted~\cite{Leitner-Ankerl2022}, whereas for the multi-threaded version,
a Cuckoo hash table is adopted~\cite{Fan2013, Li2014}. In both hash
tables, the keys are the binary representations of the determinants. Given
a key corresponding to a determinant with index $i$, the bucket of the
hash table is composed of two vectors, $B_{i,:}$ and $C_{i,:}$. Based on
our tests of \sname{}, the hash table access costs nearly half of the
runtime. Hence, in designing the algorithm and data structure of
\xsname{}, we balance the number of hash table accesses and the number of
entry updates. For each iteration in \xsname{}, where the number of hash
table accesses is the number of nonzeros in the column of $H$, we update
the entire row of $B$ and $C$, i.e., update both ground state and excited
states of the selected determinant. In \xsname{}, the hash table access
costs less than half of the runtime, and the per-iteration cost of
\xsname{} is less than $K$ times of that of \sname{}. The drawback of our
data structure implementation is that it ignores the sparsity across
states. For example, consider the scenario that for a given determinant,
the value of an excited state is non-compressible while values of other
states are all compressible. Our implementation would treat the values of
all states as non-compressible and allocate memory for them. In the
trade-off of hash table access cost and memory efficiency, we lean against
the former in the implementation of \xsname{}.

\subsubsection*{Stopping criteria}

The stopping criteria for coordinate descent methods are usually more
complicated than that for general gradient descent methods. In gradient
descent methods, the norm of the gradient is often used as the stopping
criterion. For non-stiff problems, when the norm is sufficiently small, we
are confident that the iteration is close to a first-order stationery
point. However, for coordinate descent methods, we often cannot afford to
check through the entire gradient vector, as in \xsname{}. It is also
risky to stop when the entry update $\tau \widetilde{G}_{i,:}$ is small.
Hence, in our implementation, we adopt accumulated entry updates as the
stopping criterion, i.e.,
\begin{equation} \label{eq:tol_discount}
    \mathrm{tol}_\ell = \sum_{\ell = 1}^{n} \beta^{n - \ell}
    \norm{\tau^\itell \widetilde{G}_{i^\itell,:}},
\end{equation}
where $n$ is the current iteration index, $\beta$ is a discounting factor
strictly smaller than one, and $\tau^\itell$ is the best stepsize at
$\ell$-th iteration. The accumulated entry updates could be evaluated
iteratively,
\begin{equation}
    \mathrm{tol}_\ell = \norm{\tau^{(n)} \widetilde{G}_{i^{(n)},:}}
    + \beta \cdot \mathrm{tol}_{\ell-1},
\end{equation}
and only a single $\mathrm{tol}$ needs to be kept in memory. Throughout,
the discounting factor $\beta$ is left as a hyperparameter. Given a
$\beta$, we could calculate all discounting coefficients in
\eqref{eq:tol_discount} and estimate the number of entry updates whose
coefficient is greater than $0.1$. Specifically, there are about
$-\frac{1}{\log_{10}\beta}$ entry updates with coefficients greater than
$0.1$. The suggested value for $\beta$ would be in the range of $[0.99,
0.999]$ such that about a few hundred to a few thousand entry updates are
accumulated with coefficients of the same ordering.

\section{Numerical Results}
\label{sec:num}

In this section, we perform a sequence of numerical experiments for
\ch{H2O}, \ch{C2}, and \ch{N2} under the cc-pVDZ basis set. In all
experiments, the one-body and two-body integrals are calculated by
Psi4~\cite{smith2020psi4}. The FCI excited states are calculated by our
homebrewed package CDFCI~\cite{CDFCI}. All energies are reported in
Hartree (Ha).

\subsection{\ch{H2O} excited states}

This section calculates the excited states of \ch{H2O} at equilibrium
geometry. The \ch{OH} bonds are of length 0.9751 {\r A}, and the \ch{HOH}
bond angle is $110.565^{\circ}$. The maximum memory for the \sname{}
calculation is 480 GB and the compression tolerance is 0 (no compression).
With the cc-pVDZ basis set, there are 10 electrons and 24 orbitals
involved in the calculation. Throughout, the reference energy of the
ground state is $-76.2418601$ Ha, and reference energies of excited states
are numerical results at one hundred million iterations of \xsname{}.
Reference values are attached in Appendix~\ref{app:refenergy}.

\begin{table*}[htb]
    \centering
    \begin{tabular}{ccccc}
        \toprule
        \multirow{2}{*}{Energy (Ha)} & \multicolumn{4}{c}{Number of Iterations}\\
        \cmidrule(lr){2-5}
        & $10^4$ & $10^7$ & $2 \cdot 10^7$ & $5 \cdot 10^7$ \\
        \toprule
        Ground State      & -76.2{\it 312241} & -76.24185{\it 69}
                          & -76.24185{\it 94} & -76.241860{\it 0} \\
        1st Excited State & -75.8{\it 803222} & -75.89433{\it 36}
                          & -75.89433{\it 64} & -75.8943371 \\
        2nd Excited State & -75.8{\it 452281} & -75.86048{\it 22}
                          & -75.86048{\it 51} & -75.860485{\it 8} \\
        3rd Excited State & -75.6{\it 550559} & -75.67311{\it 55}
                          & -75.67311{\it 87} & -75.673119{\it 5} \\
        4th Excited State & -75.5{\it 669476} & -75.58467{\it 40}
                          & -75.58467{\it 75} & -75.584678{\it 3} \\
        5th Excited State & -75.{\it 3466894} & -75.4844{\it 768}
                          & -75.48448{\it 24} & -75.484483{\it 6} \\
        \toprule
        Wall time (sec)   & 67.14 & 19414.28 & 38477.38 & 90759.04 \\
        \bottomrule
    \end{tabular}
    \caption{Convergence of energy of \ch{H2O}. Italics indicate
    inaccurate digits.}
    \label{tab:num-h2o}
\end{table*}

\begin{figure*}[htb]
    \begin{minipage}{0.48\textwidth}
        \centering
        \includegraphics[width=\textwidth]{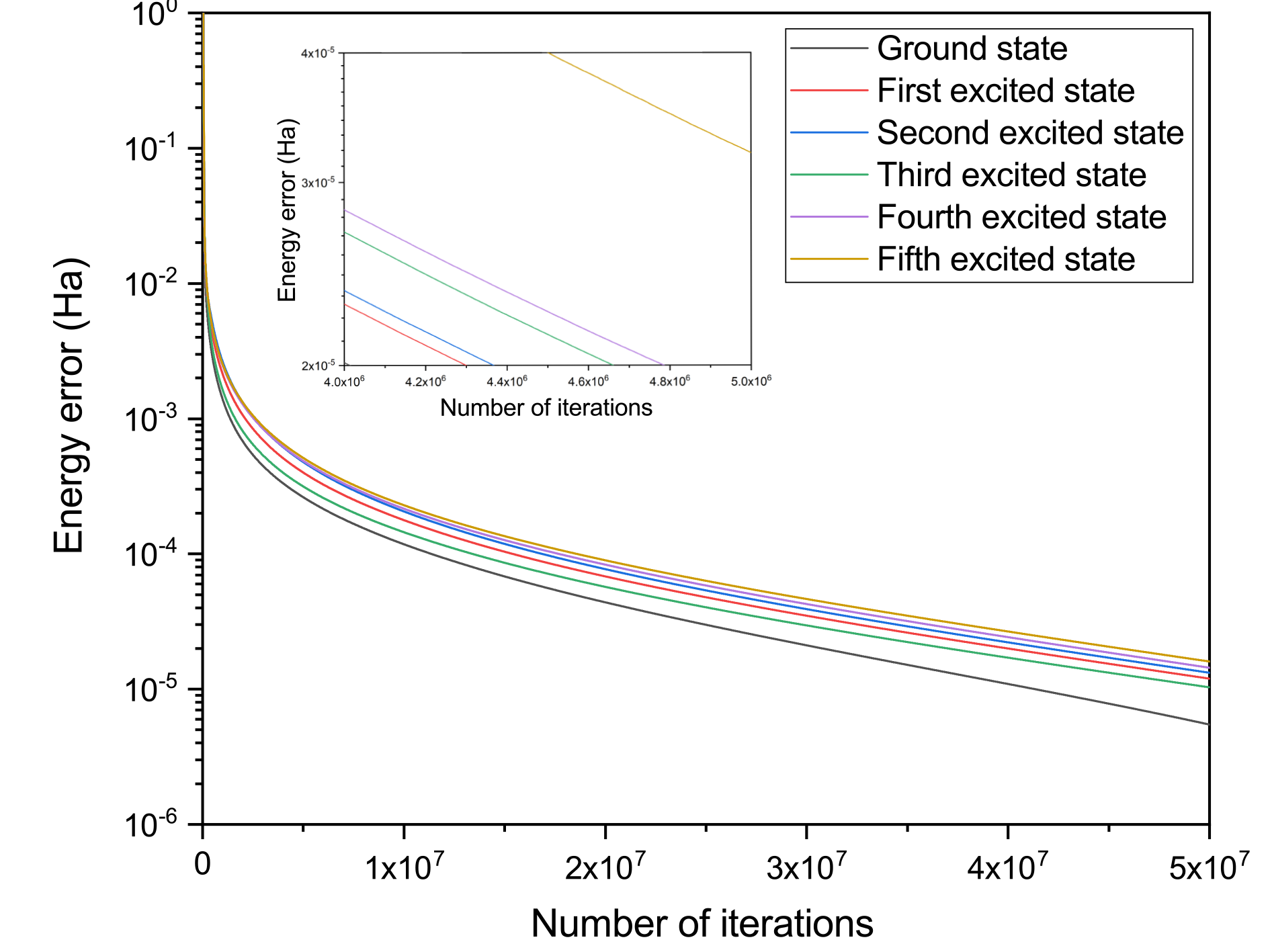}
        \caption{Convergence of energies of six low-lying excited states of
        \ch{H2O} against the number of iterations.} 
        \label{fig:num-h20}
    \end{minipage}
    ~~
    \begin{minipage}{0.48\textwidth}
        \centering
        \includegraphics[width=\textwidth]{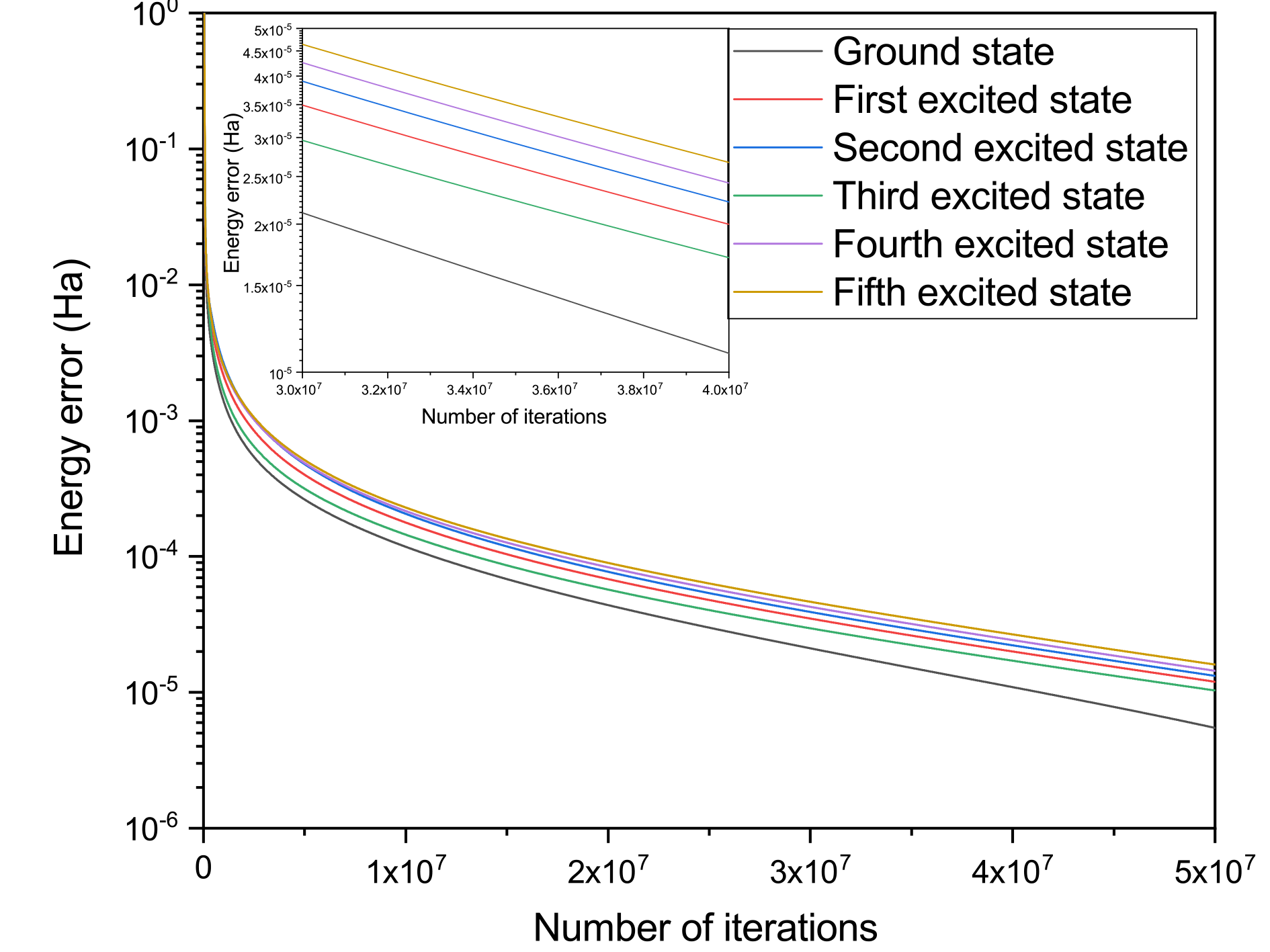}
        \caption{Convergence of energies of six low-lying excited states
        of \ch{N2} against the number of iterations. The
        threshold is $10^{-5}$.}
        \label{fig:num-n2}
    \end{minipage}
\end{figure*}

From Table~\ref{tab:num-h2o} and Figure~\ref{fig:num-h20}, we shall see
that the energy error drops quickly to the level of $10^{-4}$ mHa accuracy
at the beginning. It then has a slower but steady decay. According to
Figure~\ref{fig:num-h20}, in general, energies associated with lower
excited states are of better accuracy. The only exception for \ch{H2O} is
the energy associated with the third excited state, which achieves better
accuracy than the first and second excited state energies. From
Table~\ref{tab:num-h2o}, we find that each state can quickly converge to
the chemical accuracy. After a burn-in stage (first few thousand
iterations), the runtime is linear with respect to the number of
iterations. Hence if Figure~\ref{fig:num-h20} is redone for energy errors
against the runtime, the curves would behave similarly and the decays
remain linear against the runtime after the burn-in stage.

\subsection{\ch{N2} excited states}

This section calculates the excited states of \ch{N2} at equilibrium
geometry. Nitrogen dimer \ch{N2} is more challenging than \ch{H2O} because
the FCI problem size is much larger, so we use thresholds $10^{-4}$ and
$10^{-5}$ for compression. The \ch{N2} molecule is with bond length
1.12079 $ \r A$. The maximum memory in this section is limited to 960 GB.
With the cc-pVDZ basis set, there are 14 electrons and 28 orbitals. The
results of \ch{N2} are reported in Table~\ref{tab:num-n2-4},
Table~\ref{tab:num-n2-5} and Figure~\ref{fig:num-n2}. Throughout, the
reference energy of the ground state is $-109.28210$ Ha, and reference
energies of excited states are numerical results of \xsname{} at one
hundred million iterations. Reference values are attached in
Appendix~\ref{app:refenergy}.

\begin{table*}[htb]
    \centering
    \begin{tabular}{ccccc}
        \toprule
        \multirow{2}{*}{Energy (Ha)} & \multicolumn{4}{c}{Number of Iterations}\\
        \cmidrule(lr){2-5}
        & $10^5$ & $10^6$ & $10^7$ & $5 \cdot 10^7$ \\
        \toprule
        Ground State      & -109.2{\it 6880} & -109.28{\it 079}
                          & -109.282{\it 02} & -109.2820{\it 5} \\
        1st Excited State & -108.7{\it 1630} & -108.73{\it 197}
                          & -108.733{\it 88} & -108.733{\it 93} \\
        2nd Excited State & -108.6{\it 4476} & -108.66{\it 052}
                          & -108.662{\it 88} & -108.662{\it 94} \\
        3rd Excited State & -108.6{\it 3841} & -108.65{\it 936}
                          & -108.660{\it 81} & -108.660{\it 85} \\
        4th Excited State & -108.6{\it 0955} & -108.62{\it 886}
                          & -108.631{\it 05} & -108.631{\it 10} \\
        5th Excited State & -108.5{\it 8148} & -108.60{\it 142}
                          & -108.603{\it 65} & -108.603{\it 72} \\
        \toprule
        Wall time (sec)   & 227.8 & 1012.46 & 6671.83 & 31974.3 \\
        \bottomrule
    \end{tabular}
    \caption{Convergence of energy of \ch{N2} with threshold
    $10^{-4}$.}
    \label{tab:num-n2-4}
\end{table*}

\begin{table*}[htb]
    \centering
    \begin{tabular}{ccccc}
        \toprule
        \multirow{2}{*}{Energy (Ha)} & \multicolumn{4}{c}{Number of Iterations}\\
        \cmidrule(lr){2-5}
        & $10^5$ & $10^6$ & $10^7$ & $5 \cdot 10^7$ \\
        \toprule
        Ground State      & -109.2{\it 6836} & -109.28{\it 077}
                          & -109.282{\it 04} & -109.2821{\it 5} \\
        1st Excited State & -108.7{\it 1546} & -108.73{\it 196}
                          & -108.733{\it 90} & -108.7340{\it 7} \\
        2nd Excited State & -108.6{\it 4376} & -108.66{\it 050}
                          & -108.662{\it 91} & -108.6631{\it 1} \\
        3rd Excited State & -108.6{\it 3613} & -108.65{\it 935}
                          & -108.660{\it 83} & -108.6609{\it 6} \\
        4th Excited State & -108.6{\it 0848} & -108.62{\it 885}
                          & -108.631{\it 10} & -108.6313{\it 0} \\
        5th Excited State & -108.5{\it 8040} & -108.60{\it 141}
                          & -108.603{\it 70} & -108.6039{\it 2} \\
        \toprule
        Wall time (sec)   & 315.75 & 2140.65 & 15129.69 & 57403.06 \\
        \bottomrule
    \end{tabular}
    \caption{Convergence of energy of \ch{N2} with threshold
    $10^{-5}$.}
    \label{tab:num-n2-5}
\end{table*}

The convergence trend of \ch{N2} is similar to that of \ch{H2O} except
that the convergence rate in \ch{N2} is slower. Similarly, after the first
million iterations, \xsname{} converges linearly and the convergence rates
are quite stable for both the ground state and excited states. Therefore,
we conclude that \xsname{} is stable and efficient for various chemistry
systems with different correlation strengths. For \ch{N2}, \xsname{} takes
about ten thousand seconds to achieve chemical accuracy. Convergence rates
of all states are approximately the same. Unlike \ch{H2O}, where the
runtime scales linearly with respect to the number of iterations, for
\ch{N2}, the runtime scales sublinearly. This is mainly due to the
compression. When the compression criterium is activated, the
computational cost for compressed determinants is far less than that of
uncompressed ones. Comparing Table~\ref{tab:num-h2o} and
Table~\ref{tab:num-n2-5}, we notice that the runtime of \ch{N2} is smaller
than that of \ch{H2O}. Although the computational system of \ch{N2} is
larger, the compression with tolerance $10^{-5}$ reduces a lot of
computations and the runtime is also reduced. Comparing
Table~\ref{tab:num-n2-4} and Table~\ref{tab:num-n2-5}, we find that the
accuracies for both ground state and excited state energies are at the
same level of the truncation threshold. When a smaller truncation
threshold is used, the runtime is longer whereas the accuracies are
consistently improved. Therefore, the compression technique is efficient
and reliable.

\subsection{Carbon dimer binding curves}

In this section we test \ch{C2} with bond lengths form $1 \r A$ to $2.6 \r
A$. We computed five low-lying energies of singlet of \ch{C2}. The
symmetry in the basis set is implemented via the Hartree-Fock calculation,
i.e., in the Psi4 calculation. More precisely, the singlet calculation is
realized by setting the molecule as a singlet and its irreducible
representations. The maximum memory in this section is 120 GB and the
tolerance is 0. With the cc-pVDZ basis set, there are 12 electrons and 56
orbitals. We perform 1 million iterations for \xsname{}. In all
configurations, the accuracies for all states are at the level of chemical
accuracy.

\begin{table*}[htb]
    \centering
    \begin{tabular}{cccccc}
    \hline
    \multirow{2}{*}{R(\r A)} &
    \multicolumn{5}{c}{Energy of five low-lying states (Ha)} \\
    \cline{2-6}
    &0th &1st &2nd &3rd &4th \\ \hline
    1.0&-75.55231& -75.37074& -75.34005& -75.25824& -75.24635\\
    1.1&-75.67528& -75.52584& -75.52314& -75.42099& -75.40454\\
    1.2&-75.7246& -75.6188& -75.61144& -75.51174& -75.46344\\
    1.3&-75.73152& -75.66195& -75.65091& -75.55151& -75.4995\\
    1.4&-75.71569& -75.67459& -75.66213& -75.56135& -75.5052\\
    1.5&-75.68951& -75.67034& -75.65703& -75.55471& -75.49432\\
    1.6&-75.66102& -75.65712& -75.64203& -75.53995& -75.47353\\
    1.7&-75.64014& -75.63676& -75.62022& -75.52375& -75.4532\\
    1.8&-75.62201& -75.6169& -75.59619& -75.5117& -75.45362\\
    1.9&-75.60453& -75.59944& -75.57506& -75.50693& -75.44587\\
    2.2&-75.56245& -75.55941& -75.53746& -75.51137& -75.46051\\
    2.5&-75.53929& -75.53814& -75.52593& -75.51654& -75.49545\\ \hline
    \end{tabular}
    \caption{Energy of five low-lying states of \ch{C2} in singlet.}
    \label{tab:num-bind-c2-singlet}
\end{table*}

\begin{figure}
    \centering
    \includegraphics[width=0.5\textwidth]{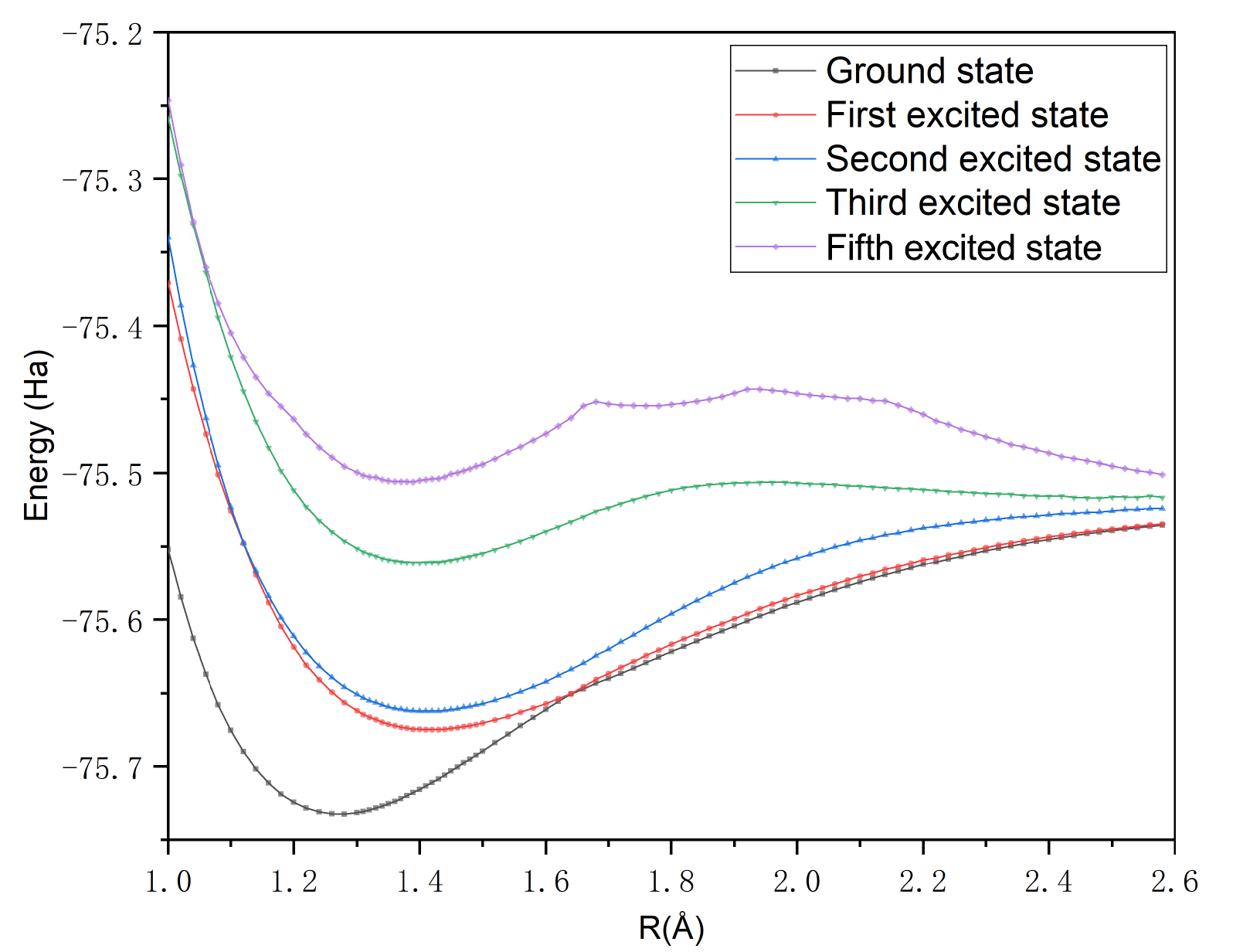}
    \caption{Low-lying potential energy surfaces of carbon dimer in
    singlet the cc-pVDZ basis.}
    \label{fig:num-bind-c2-singlet}
\end{figure}

The energies of five low-lying states of \ch{C2} in singlet
$({\prescript{1}{}{\Sigma}_g})$ are shown in
Table~\ref{tab:num-bind-c2-singlet}. Binding curves are depicted in
Figure~\ref{fig:num-bind-c2-singlet}. In general, we observe that the
binding curves for lower energy states are smoother in
Figure~\ref{fig:num-bind-c2-singlet}. We find a lot of cross-over points.
Each cross-over point corresponds to a configuration whose energies are
degenerate. Lower energy binding curves have fewer cross-over points. The
binding curve for the fourth excited state has many cross-over points with
binding curves of higher excited states though they are not calculated.

\section{Conclusion and Discussion}
\label{sec:conclusion}

We proposed \xsname{} in this paper as an efficient low-lying excited
states solver under the FCI framework. \xsname{} adopts an extension of
the objective function in the \sname{} method. More precisely, \xsname{}
extends the single-column version (ground state) to a multi-column version
(low-lying excited states) and leads to \eqref{eq:opt}. Then a tailored
coordinate descent method is applied to address~\eqref{eq:opt}. \xsname{}
first selected a determinant with the largest entry in magnitude in the
approximated gradient, and then the selected row of the iteration variable
$C$ is updated, i.e., the coefficients of a determinant for all states are
updated. To avoid memory overflow, a hard-thresholding type compression is
applied to $B \approx H C$ for $H$ being the Hamiltonian matrix, which in
turn limits the growth of nonzeros in $C$. Finally, we carefully maintain
the double precision accuracy of $C^\top C$ and $C^\top H C = C^\top B$,
and estimate the eigenvalues through a generalized Rayleigh quotient
procedure. Based on results from the theory of numerical
analysis~\cite{Golub2013}, the ground state and low-lying excited states
are of the first order accuracy, whereas the ground state energy and
excited state energies are of the second order accuracy. In summary,
\xsname{} extends \sname{} to calculating low-lying excited states and
inherits almost all desired properties of \sname{}. Numerical results on
various chemistry systems demonstrate the efficiency of \xsname{}.

Extending \sname{} to higher-lying excited states is feasible but more
challenging. Memory cost is a major concern. When more excited states are
computed using \sname{}, the number of columns in $C$ and $B$ is
increased. At the same time, the number of nonzero rows in $C$ and $B$
also needs to be increased to incorporate the sparsity of higher-lying
excited states. Hence, the memory increases faster than linear scaling
with respect to the number of excited states. Besides the memory cost, the
degeneracy in higher-lying excited states would also cause trouble if $K$
is not properly chosen.

There are a few promising future directions. First of all, \xsname{} has
not fully exploited the sparsity of the low-lying excited states. Due to
the nature of \eqref{eq:opt}, the objective function is rotation
invariant, i.e., the objective function remains the same for $C$ and $CQ$
with $Q$ being an orthogonal matrix. Hence, \xsname{} can converge to the
eigenspace formed by desired ground state and low-lying excited states.
While it is not guaranteed to converge to the sparse eigenvectors
directly. Some recent works~\cite{Gao2022, Gao2023a, Gao2023b} provide
promising paths to address the sparsity issue. Second, the basis sets, so
far, remain the Hartree Fock molecular orbitals. Applying orbital
optimization methods like CASSCF~\cite{Olsen2011} or
OptOrbFCI~\cite{Li2020} with state-averaged idea together with \xsname{}
would be a direct extension. While exploring various orbital rotations for
different excited states coupled with \xsname{} would be an interesting
future direction. Lastly, we did not fully incorporate the compressed
evaluation of the Hamiltonian matrix and other perturbative approximations
as in other FCI excited state work~\cite{Holmes2017, Greene2022}, which
could be combined with \xsname{} to further accelerate the proposed
method.

 
\begin{acknowledgement}
    The work of ZW is supported by the US National Science Foundation
    under awards DMS-1454939 and OAC-1450280. This work is part of ZW's
    PhD thesis at Duke University. YL is supported in part by National
    Natural Science Foundation of China (12271109) and Shanghai Pilot
    Program for Basic Research - FuDan University 21TQ1400100 (22TQ017).
\end{acknowledgement}







\bibliography{reference}

\newpage
\appendix

\section{Optimal Stepsize via Linesearch}
\label{app:linesearch}

The optimal stepsize $\tau$ could be obtained by solving
\eqref{eq:linesearch}. The function $f(C^\itell + \tilde{\tau}
e_{i^\itellp} \widetilde{G}_{i^\itellp, :})$ could be rewritten as a
forth-order polynomial of $\tilde{\tau}$. For the sake of notation, we
omit all superscripts of the iteration index and obtain, 
\begin{equation} \label{eq:4thpoly}
    f(C + \tilde{\tau} e_i
    \widetilde{G}_{i,:})
    = c_0 + c_1 \tilde{\tau} + c_2 \tilde{\tau}^2
    + c_3 \tilde{\tau}^3 + c_4 \tilde{\tau}^4,
\end{equation}
where the polynomial coefficients are,
\begin{align}
    c_0 ={}& f(C), \\
    c_1 ={}& \norm{ \widetilde{G}_{i,:}}^2,\\
    \begin{split}
        c_2 ={}&
        2 H_{i,i} \norm{\widetilde{G}_{i,:}}^2
        + 2 \widetilde{G}_{i,:} \big( C^\top
        C\big) \widetilde{G}_{i,:}^\top \\
        & + 2 \big( C_{i,:} \widetilde{G}_{i,:}^\top \big)^2
        + 2 \norm{C_{i,:}}^2 \norm{\widetilde{G}_{i,:}}^2,
    \end{split} \\
    c_3 ={}&
    4 \big( C_{i,:} \widetilde{G}_{i,:}^\top \big)
    \norm{\widetilde{G}_{i,:}}^2, \\
    c_4 ={}& \norm{\widetilde{G}_{i,:}}^4. 
\end{align}
Notice that coefficient $c_2$ could be evaluated in $O(K^2)$ operations,
and coefficient $c_1$, $c_3$ and $c_4$ could be evaluated in $O(K)$
operations, where $K$ is the number of states and length of all row
vectors. 

Finding the minimum of the forth-order polynomial could be addressed via
solving a third-order polynomial, 
\begin{equation} \label{eq:3rdpoly}
    c_1 + 2c_2 \tilde{\tau} + 3c_3 \tilde{\tau}^2 + 4c_4 \tilde{\tau}^3
    = 0.
\end{equation}
There are three scenarios in solving \eqref{eq:3rdpoly}: 1) one root; 2)
two roots; 3) three roots. When there is only one root, it achieves the
minimum of \eqref{eq:4thpoly}. When there are two roots, one of which is
of multiplicity one and it achieves the minimum. When there are three
roots, the one further away from the middle one achieves the minimum.
Through the above procedure, the linesearch problem \eqref{eq:linesearch}
could be addressed efficiently in $O(K^2)$ operations.

\section{Rayleigh Quotient}
\label{app:rayleigh}

Let $H$ be a symmetric matrix of size $N$. The eigenvalues of $H$ are
denoted as $E_0 < E_1 < \cdots < E_{N-1}$. And the associated eigenvectors
are $V_0, V_1, \dots, V_{N-1}$. For simplicity, we assume that $H$ is a
gapped matrix. Given a vector $x \in \bbR^N$, the Rayleigh quotient is
defined as,
\begin{equation}
    r(x) = \frac{x^\top H x}{x^\top x}.
\end{equation}
Obviously, the Rayleigh quotient is $x$ scale-invariant, i.e., $r(x) =
r(\alpha x)$ for any nonzero scaler $\alpha$. Hence, we could focus on a
normalized vector $x$ such that $\norm{x} = 1$.

An interesting and useful property of the Rayleigh quotient is that $r(x)$
is a quadratically accurate estimate of an eigenvalue. More precisely, let
$V_j$ be one of the eigenvectors of $H$. We consider the case that $x$ is
sufficiently close to $V_j$, i.e., $\norm{x - V_j} = O(\epsilon)$ for
$\epsilon$ small. Then an important consequence of the Rayleigh quotient
is that~\cite{RN1158}
\begin{equation}
    r(x) = r(V_j) + O(\epsilon^2) = E_j + O(\epsilon^2).
\end{equation}
This is the second-order accuracy we are referring to in the main paper.

In this paper, instead of the Rayleigh quotient of a single vector, we
adopt a generalized Rayleigh quotient (or block Rayleigh quotient), as in
\eqref{eq:geneig}. The eigenvalue estimation is via solving a generalized
eigenvalue problem. We could view the generalized eigenvalue problem step
as a normalization step so that each column in $CU$ as in \eqref{eq:VCU}
is a normalized and aligned estimation of an eigenvector of $H$. Then the
quadratically accurate property of the Rayleigh quotient remains valid in
the generalized Rayleigh quotient case.

\section{Reference Energies}
\label{app:refenergy}

The reference energies for \ch{H2O} and \ch{N2} under cc-pVDZ basis are
reported in Table~\ref{tab:ref-energy-h2o-n2}.

\begin{table}[htb]
    \centering
    \begin{tabular}{ccc}
        \toprule
        & \ch{H2O} & \ch{N2} \\
        \toprule
        0th ES & -76.241860063 & -109.282165 \\
        1st ES & -75.894337144 & -108.734087 \\
        2nd ES & -75.860485864 & -108.663124 \\
        3rd ES & -75.673119564 & -108.660977 \\
        4th ES & -75.584678392 & -108.631318 \\
        5th ES & -75.484483689 & -108.603936 \\
        \bottomrule
    \end{tabular}
    \caption{Reference energies (Ha) for \ch{H2O} and \ch{N2} under cc-pVDZ
    basis.}
    \label{tab:ref-energy-h2o-n2}
\end{table}

\end{document}